\documentclass[letterpaper,11pt,leqno]{article}
\usepackage[utf8]{inputenc}
\usepackage{paper}
\usepackage{threeparttable}  
\usepackage{booktabs}        
\usepackage[hyphens]{url}
\usepackage{hyperref}
\usepackage{booktabs}
\usepackage{threeparttable}
\usepackage{adjustbox}
\usepackage{pdflscape}
\usepackage{longtable}
\usepackage{makecell}
\usepackage{graphicx}
\usepackage{float}
\hypersetup{pdftitle={Minimalist LaTeX Template for Academic Papers}}

\newcommand{\bib}{paper.bib}

\begin{document}

\title{Recession Detection Using Real Time GDP Data}

\author{Neha Sikand and Rongjin Zhang
%
\thanks{Department of Economics, University of California, Santa Cruz. \\We are deeply indebted to Pascal Michaillat for his invaluable advice, continuous guidance, encouragement, and support.}}

\date{May 2026}   


\begin{titlepage}
\maketitle


\begin{abstract}


This paper examines whether real-time GDP announcements can reliably identify business-cycle turning points. Using U.S. real-time GDP vintages from 1947 to 2021, we construct 4,356 recession indicators based on alternative smoothing methods and scaling variations. We then combine these indicators with alternative thresholds to generate 137,457 perfect recession classifiers. The selected classifiers identify all 12 historical recessions without generating false positives or false negatives. Restricting attention to the high-precision segment yields two classifiers with a standard deviation of detection errors below three months, while the selected ensemble signals recessions, on average, 3.04 months after their official onset. The framework accurately identifies recession episodes across vintages, suggesting that discrepancies in prior work may reflect limitations of traditional dating methods in addition to data revisions. Overall, the results indicate that real-time GDP announcements provide a practical proxy for NBER-style recession dating.

\end{abstract}

\end{titlepage}

\section{Introduction}\label{s:introduction}

Business-cycle turning points play a central role in macroeconomic analysis and policy decisions, yet recession dating remains difficult in real time and lacks formal institutional recognition in many countries. While the United States relies on the National Bureau of Economic Research to identify recessions, many economies either announce turning points with substantial delays or rely on simplified heuristic rules such as two consecutive quarters of negative GDP growth. This paper asks whether real time GDP announcements themselves can serve as a reliable measure of business-cycle turning points.

Building on \cite{michaillat2025}, this paper answers this question by evaluating the performance of real time GDP against official recession dates determined by the National Bureau of Economic Research (NBER). We construct 4,356 recession indicators using real time GDP vintages with alternative smoothing methods and scaling variation. We then combine these indicators with different thresholds to obtain our classifiers. The selected 137,457 classifiers are statistically perfect in the sense that they identify all 12 historical recessions in the 1947–2021 training sample without generating false positives. Among these, a subset of classifiers lies on the anticipation–precision frontier. Restricting attention to the high-precision segment yields two classifiers with a standard deviation of detection errors below three months. The selected ensemble signals recessions, on average, 3.04 months after their true onset.

This paper contributes to the literature on real time business-cycle dating by showing that real time GDP announcements provide a reliable basis for recession dating across vintages. This suggests that discrepancies in prior work may reflect limitations of traditional dating methods in addition to data revisions. Overall, the results suggest that real time GDP announcements contain substantial information about business-cycle turning points and may provide a practical real time proxy for NBER-style recession dating, particularly for countries without formal business-cycle dating committees.




\section{Related Literature}
More broadly, this paper bridges the literature on real time GDP measurement and the literature on recession-classification methods. While previous work studies either GDP-based recession dating or industrial-production-based classification algorithms separately, this paper shows that real time GDP releases can support timely and stable recession detection within a recession-classification framework.

\cite{hamilton2011} shows that real time GDP data contain meaningful information about business-cycle turning points, although recession dating remains difficult because initial GDP releases are noisy and subject to substantial revisions. Similarly, \cite{crumpGiannoneLucca2020} show that business-cycle dating based on real GDP can differ considerably across data vintages, with some recessions appearing or disappearing as GDP estimates are revised over time. Together, these findings highlight the importance of evaluating recession indicators using real time vintage data rather than ex post revised series. Motivated by this literature, this paper evaluates whether real time GDP announcements provide timely and reliable signals of recession turning points.

Building on this literature, \cite{michaillat2025} evaluates whether recession-classification methods can improve real time recession detection using industrial production data. Although the National Bureau of Economic Research relies heavily on product-market variables to date business cycles, the paper finds that industrial-production classifiers perform worse than labor-market classifiers in terms of both anticipation and precision. These findings suggest that not all output-based indicators are equally informative for recession detection and motivate this paper’s examination of whether broader output measures such as real time GDP announcements can support more timely and stable recession detection.

\section{Data}\label{s:introduction}
This section presents the data on business cycle dating and real time GDP announcements used in the paper. The sample covers the period from January 1947 to December 2025.
\subsection{Real Time GDP Data}
This paper uses real time macroeconomic data from the Federal Reserve Bank of Philadelphia’s real time Data Set for Macroeconomists (RTDSM), which records National Income and Product Accounts (NIPA) variables as they were available to policymakers and researchers at each point in time. The data consist of historical “vintages” that preserve both initial releases and subsequent revisions, allowing the analysis to replicate the information available in real time rather than relying on fully revised data. The vintages are constructed using publications from the Bureau of Economic Analysis (BEA), including the \textit{Survey of Current Business} and historical NIPA tables. Quarterly vintages begin in 1947 and generally contain the BEA’s advance estimate for the previous quarter, while monthly vintages track the sequence of advance, second, and third estimates. This structure makes it possible to evaluate recession detection using only information that would have been available to policymakers in real time.

\subsection{Business Cycle Dating}
\citet{nber2023} identifies US recessions through
its Business Cycle Dating Committee. The NBER does not identify recessions using a single
economic indicator. Instead, it evaluates a broad range of macroeconomic variables to
determine turning points in the business cycle. A business cycle peak marks the end
of an economic expansion and the beginning of a contraction, while a trough marks
the end of the contraction and the start of a recovery (NBER, 2004) \nocite{nber2024}. Following the NBER
convention, the first month of a recession is defined as the month after the peak, and the
last month is defined as the month of the trough. NBER recession dates play an important
role in macroeconomic research because they provide the benchmark chronology used
to evaluate business cycle fluctuations and recession forecasting models. However, these
official dates are often announced with substantial delays, which limits their usefulness for
real time policymaking and motivates the development of recession detection algorithms.

Between January 1947 and December 2025, the National Bureau of Economic Research (NBER) dates 12 U.S. recessions, which are shown in Figure 1. However, recession start dates are identified only with a considerable delay: from 1979 to 2021, the NBER announced recessions an average of 6.3 months after their onset.

\section{Real Time GDP }
\begin{figure}[t!]
    \centering
    \includegraphics[width=1\linewidth]{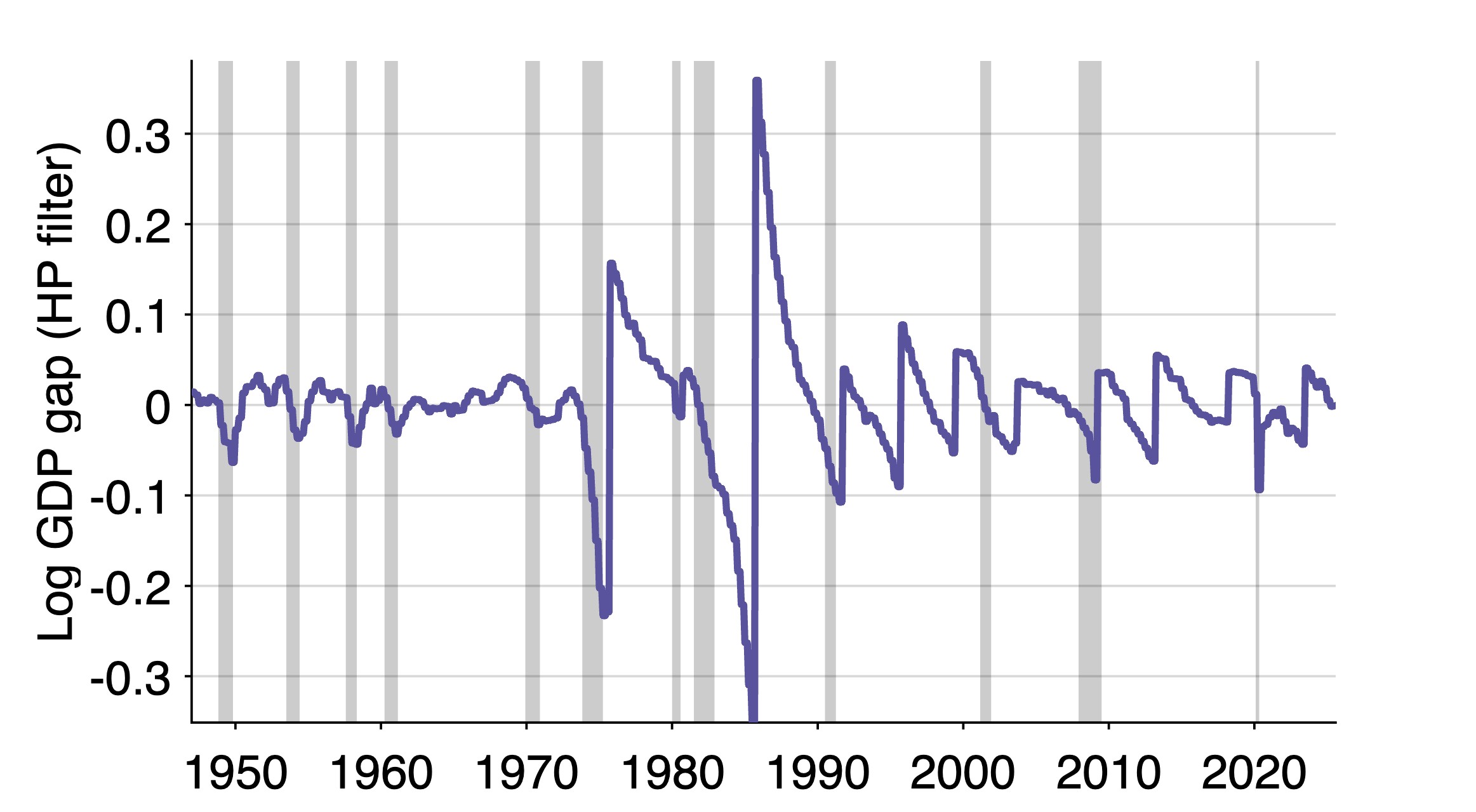}
    \caption{Real GDP First Release}
    \label{fig:figure_data}
    
    \vspace{0.3em}
    \begin{minipage}{0.95\linewidth}
    \footnotesize
    \textit{Notes:} The figure plots the real-time log GDP gap in the United States from 1947–2025, computed using a Hodrick–Prescott filter. Positive values indicate that GDP is above trend, while negative values indicate that GDP is below trend. Gray shaded bars indicate recession periods dated by the NBER.

    \end{minipage}
\end{figure}

Figure \ref{fig:figure_data} plots real time announcements of U.S. real gross domestic product (GDP) from 1947 to 2025. The series exhibits strong long-run growth, with periods of slower growth and temporary declines during economic downturns. Because the data are based on first-release GDP announcements, the figure reflects the information that would have been available to policymakers and researchers in real time rather than revised historical estimates. Gray shaded bars indicate recessions dated by the National Bureau of Economic Research (NBER). The figure highlights how GDP growth slows markedly during recession periods, particularly during the Great Recession and the COVID-19 recession.

The real time GDP indicator is constructed in three steps: smoothing, curvature adjustment, and turning-point detection. First, quarterly real time GDP data are smoothed using backward-looking simple moving averages over windows ranging from 0 to 4 quarters in order to reduce short-run volatility. Second, the smoothed series is transformed using a Box--Cox transformation with curvature parameters ranging from 0 to 1, allowing the indicator to vary between logarithmic and linear specifications. Finally, the real time GDP indicator is constructed by comparing the transformed GDP series to its recent peak over backward-looking windows of 1 to 6 quarters. Because GDP is procyclical, the indicator measures the decline in GDP relative to its recent maximum, so larger values correspond to larger output declines and therefore a higher likelihood of recession. The procedure generates a large set of candidate real time GDP indicators that differ according to their smoothing, curvature, and turning-point parameters.

\begin{figure}[H]
    \centering
    \includegraphics[width=1\linewidth]{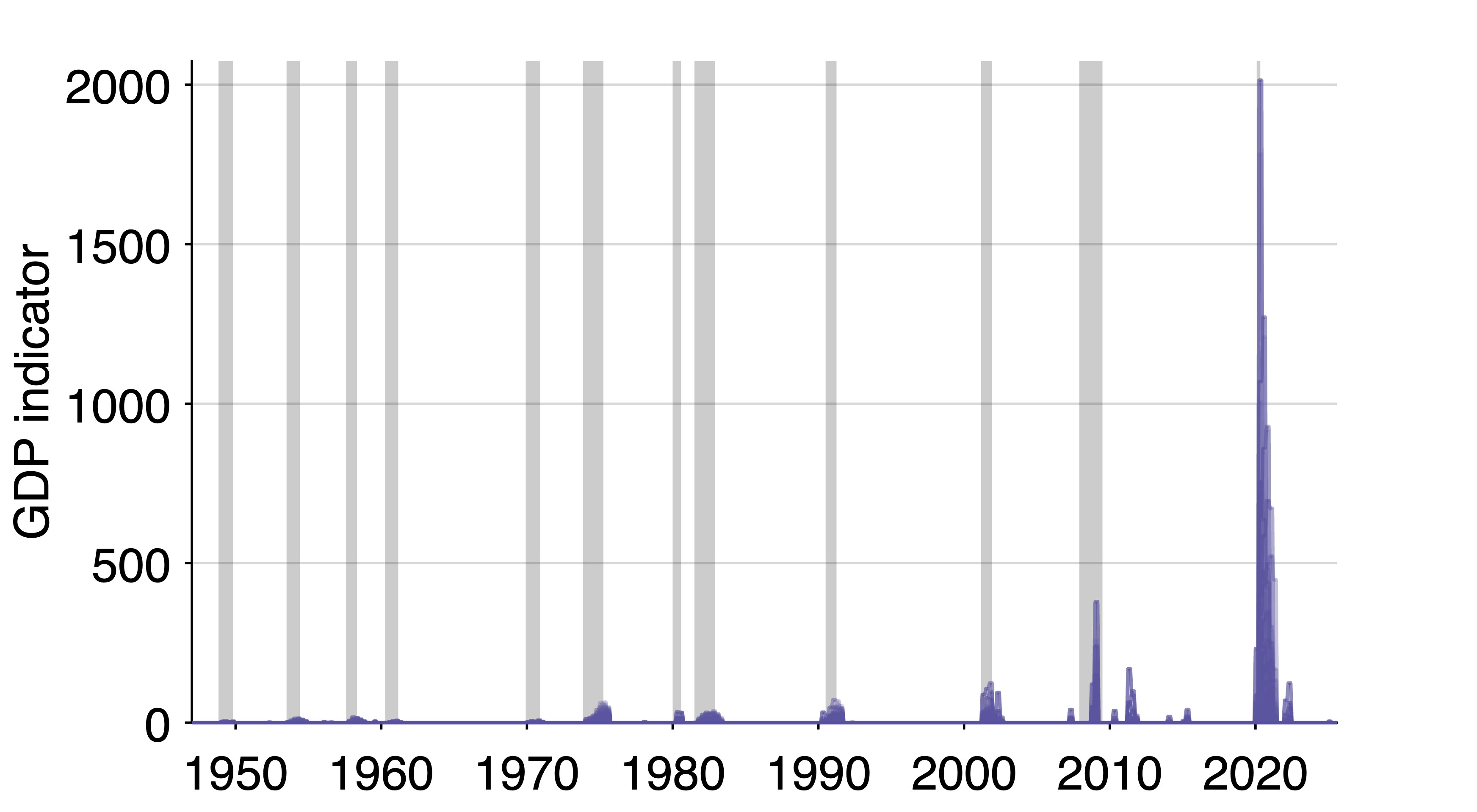}
    \caption{Real Time GDP Indicator}
    \label{fig:gdp_indicator}

    \vspace{0.3cm}
    \begin{minipage}{0.95\linewidth}
    \footnotesize
    \textit{Notes:} The figure displays the set of real time GDP recession indicators constructed from quarterly GDP vintages. Indicators are generated by combining different smoothing, curvature, and turning-point parameters. Shaded vertical bars denote NBER recessions. Larger values indicate larger declines in GDP relative to its recent peak and therefore a higher likelihood of recession.
    \end{minipage}
\end{figure}

\section{Constructing and Evaluating Recession Classifiers}
 We begin by generating a large set of candidate recession indicators and threshold combinations. Each candidate classifier is then evaluated according to its ability to reproduce official recession episodes without generating false positives or false negatives. From this set of perfect classifiers, we identify those that lie on the anticipation-precision frontier and evaluate their real-time recession detection performance. Finally, following \citet{michaillat2025}, we construct an ensemble of frontier classifiers rather than selecting a single recession rule, thereby avoiding the need to impose a specific preference over anticipation versus precision.

\subsection{Detection Methodology}

Using 4,356 recession indicators together with 2,500 alternative thresholds ranging from 0.0001 to 0.25, we generate 137,457 perfect recession classifiers. A recession begins when the indicator crosses the threshold from below while the economy is currently in expansion. To avoid false recession signals caused by temporary movements around the threshold, the methodology keeps track of whether the economy is already in a recession or expansion state. In particular, the economy can only enter a new recession if it was previously in expansion, and it only returns to expansion once the indicator falls back to zero. This prevents short-lived rebounds in the indicator during recoveries from being incorrectly treated as new recessions. Although more computationally intensive than simple threshold rules, the procedure produces a more stable and economically intuitive classification of business-cycle turning points.

\subsection{Selecting Perfect Classifiers}
For each classifier, we evaluate its ability to identify official United States recession episodes over the sample period. We retain only classifiers that correctly identify all recessions without generating false positives or false negatives. A false negative occurs when a classifier fails to detect a true recession, whereas a false positive occurs when a classifier incorrectly signals a recession during an expansion. Restricting attention to the high-precision segment, we then compare the mean and standard deviation of detection errors across classifiers to identify recession rules that deliver the strongest real time turning-point performance.

\subsection{Evaluating Perfect Classifiers}
To select the classifiers included in the ensemble, we follow \citet{michaillat2025}. In principle, a policymaker with mean--variance preferences over detection error would choose the frontier classifier $k$ that minimizes
\[
\mu(k) + \lambda \sigma(k),
\]
where $\mu(k)$ and $\sigma(k)$ denote the mean and standard deviation of the detection error, and $\lambda > 0$ captures the weight placed on precision relative to anticipation. Because $\lambda$ is not observed, we do not select a single classifier. Instead, as in \citet{michaillat2025}, we retain all classifiers on the anticipation--precision frontier whose standard deviation of detection errors satisfies
\[
\sigma(k) < 3 \text{ months}.
\]
Under the normal approximation for detection errors, this restriction implies that the associated 95\% confidence interval for the recession start date has width
\[
4\sigma(k) < 12 \text{ months}.
\]
We therefore define the ensemble as
\[
\mathcal{K}^{*} = \{ k \in \mathcal{F} : \sigma(k) < 3 \},
\]
where $\mathcal{F}$ denotes the set of classifiers on the anticipation--precision frontier. This is the same selection rule used in \citet{michaillat2025}: it focuses attention on the high-precision segment of the frontier and excludes classifiers whose implied timing uncertainty exceeds one year.

\subsection{Finding the Anticipation-Precision Frontier}
\begin{figure}[t!]
    \centering

    \begin{subfigure}[t]{0.9\linewidth}
        \centering
        \includegraphics[width=\linewidth]{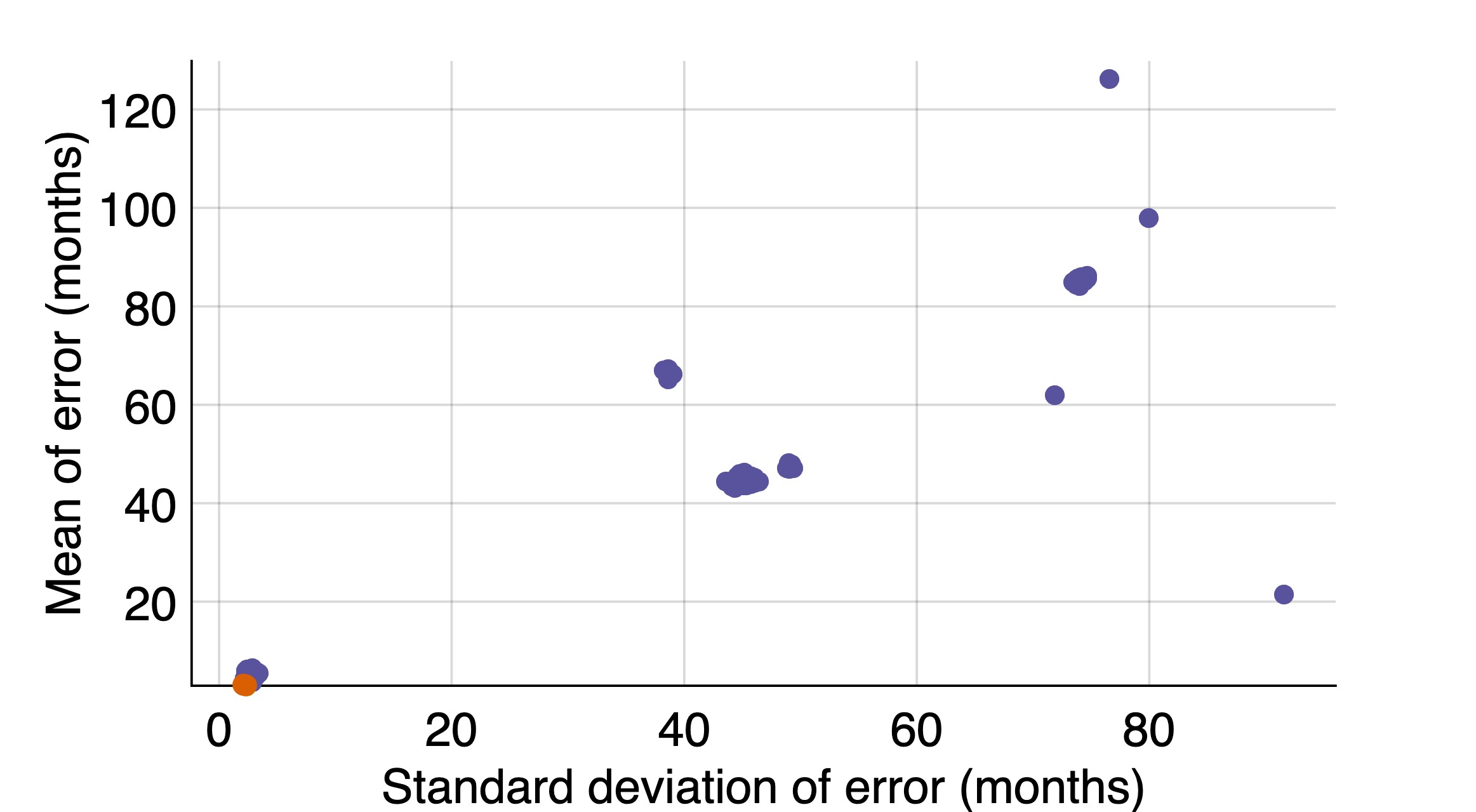}
        \caption{4 Perfect Classifiers on the Anticipation--Precision Frontier}
        \label{fig:figure_frontier}
    \end{subfigure}

    \vspace{0.4cm}

    \begin{subfigure}[t]{0.9\linewidth}
        \centering
        \includegraphics[width=\linewidth]{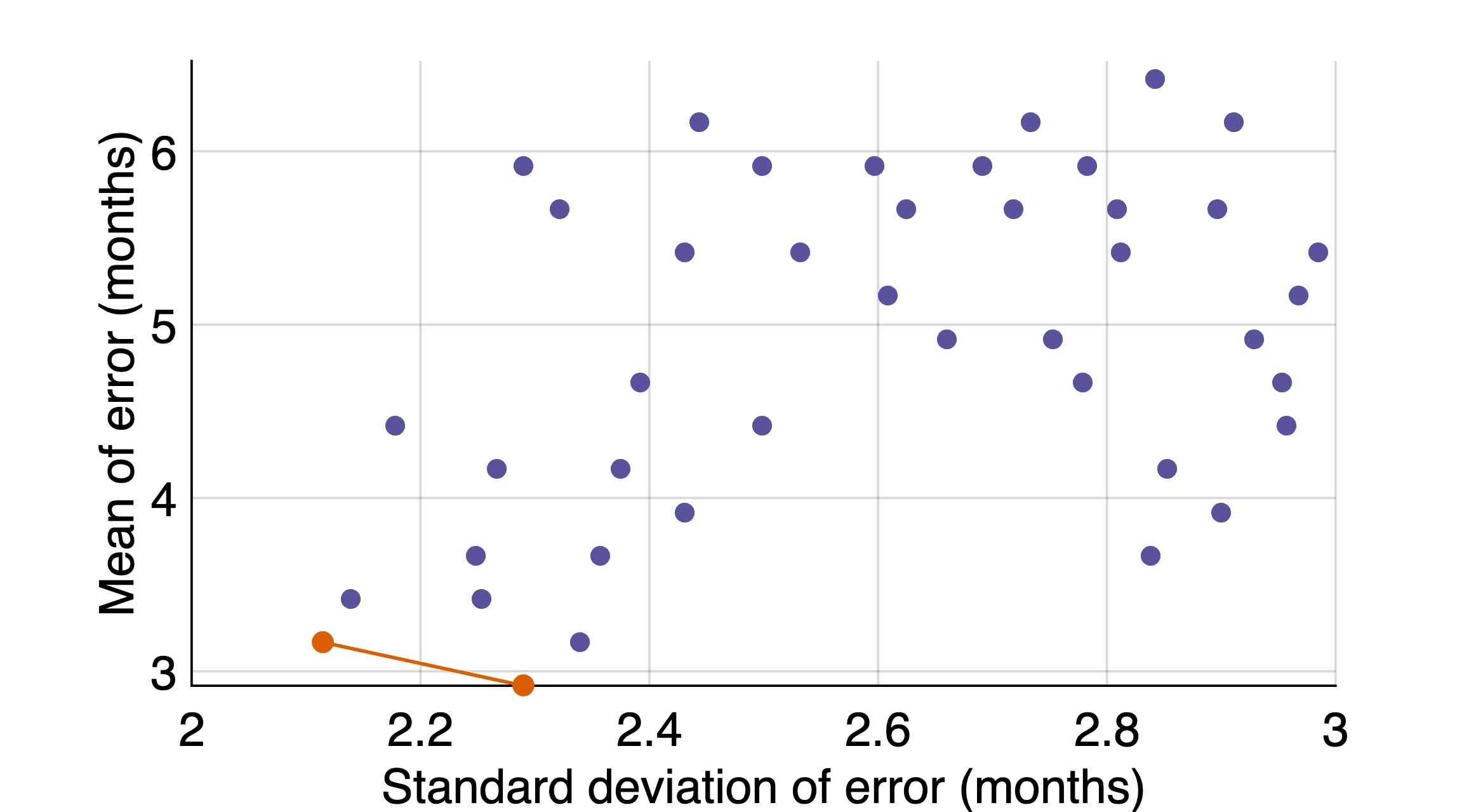}
        \caption{2 Perfect Classifiers satisfying $\sigma(k) < 3$ on the Anticipation--Precision Frontier}
        \label{fig:figure_frontier_precision}
    \end{subfigure}

    \caption{Anticipation and Precision of 137,457 Perfect Recession Classifiers in the United States, January 1947--December 2021}
    
    \label{fig:frontier_combined}

    \vspace{0.3em}
    \begin{minipage}{0.95\linewidth}
    \footnotesize
    \textit{Notes:} The figures plot the mean detection error against the standard deviation of detection errors for recession classifiers constructed using alternative indicators and thresholds. Each point represents a classifier. The orange frontier traces classifiers that maximize anticipation for a given level of dispersion, or equivalently maximize precision for a given level of anticipation. Panel (a) shows the full anticipation--precision frontier across all classifiers, while panel (b) focuses on the high-precision region satisfying $\sigma(k) < 3$. Detection errors are measured relative to recession start dates dated by the NBER.
    \end{minipage}
\end{figure}

We select classifiers along the anticipation–precision frontier, defined as those that minimize both the mean detection delay and its variability. This frontier highlights classifiers that achieve the best tradeoff between timely and reliable recession detection. Figure ~\ref{fig:figure_frontier} plots the mean and standard deviation of detection errors across a large set of recession classifiers. The anticipation–precision frontier identifies the set of optimal classifiers that minimize detection delay for a given level of dispersion, or equivalently, minimize dispersion for a given level of delay. The downward-sloping frontier reveals a clear tradeoff: classifiers that achieve greater anticipation (more negative mean errors) exhibit substantially higher variability in detection timing, while classifiers with low dispersion detect recessions more consistently but with a modest delay.

The anticipation--precision frontier contains 2 classifiers out of the 137,457 statistically perfect classifiers. The most precise classifier has a mean detection error of 3.17 months and a standard deviation of detection errors of 2.11 months. The second frontier classifier detects recessions slightly earlier, with a mean detection error of 2.92 months, but at the cost of a modest reduction in precision, as reflected by a higher standard deviation of 2.29 months. Overall, both classifiers identify recessions 3.04 months after their official start dates while maintaining relatively precise timing. For comparison, \citet{michaillat2025} finds that the selected classifier ensemble for the United States detects recessions only 1.2 months after their onset.

\subsection{Selecting the Frontier Classifiers with Highest Precision}
There is no uniquely optimal classifier along the anticipation--precision frontier because the preferred tradeoff between early detection and precision depends on policymakers’ preferences. Following \citet{michaillat2025}, we therefore focus on the high-precision segment of the frontier by retaining all classifiers with a standard deviation of detection errors below 3 months. Under the assumption of normally distributed detection errors, this restriction implies a 95\% confidence interval for recession start dates narrower than 12 months.
\begin{table}[t!]
\centering
\caption{Classifier Performance Metrics}
\footnotesize

\begin{threeparttable}

\resizebox{\textwidth}{!}{%
\begin{tabular}{lcccccc}
\toprule
\textbf{Smoothing} & \textbf{Smooth.} & $\gamma$ & $\beta$ & \textbf{Threshold} $\zeta$ & \textbf{Std. error} & \textbf{Mean error} \\
\textbf{method} & \textbf{parameter} &  &  &  &  &  \\
\midrule
SMA & 2 & 0.0 & 1 & 0.0015 & 2.11 & 3.17 \\
SMA & 2 & 0.1 & 1 & 0.0027 & 2.29 & 2.92 \\
\bottomrule
\end{tabular}
}

\begin{tablenotes}[flushleft]
\footnotesize
\item \parbox{\textwidth}{\textit{Notes:} Detection-error statistics are reported in months. \texttt{SMA} denotes simple moving average smoothing. The table reports classifiers selected from the anticipation--precision frontier subject to the restriction $\sigma(k) < 3$ months. The Box--Cox parameter $\gamma$ governs the transformation of the recession indicator, while the turning parameter $\beta$ determines the horizon used to compute turning points. A recession is identified when the resulting indicator crosses the threshold $\zeta$ from below. Detection errors are defined as the difference, in months, between estimated and official recession start dates dated by the NBER.}
\end{tablenotes}

\end{threeparttable}

\label{tab:classifier-ensemble}
\end{table}




\section{Detecting Recessions}

To build the recession detection algorithm, we follow \cite{michaillat2025} to aggregate the detection signals produced by the classifier ensemble. We then use the aggregated signals to compute the current recession risk. 

\subsection{Empirical Construction of Recession Probabilities}

Using the methodology of \citet{michaillat2025}, we compute the detection error for each classifier \(k\) and recession \(j\) in the training sample.
\[
\varepsilon_{k,j} = d_{k,j} - s_j,
\]

where $d_{k,j}$ is the classifier's detection date and $s_j$ is the NBER recession start date. For each classifier, these historical errors are summarized by their sample mean $\mu_k$ and standard deviation $\sigma_k$, which capture systematic anticipation or delay and timing precision, respectively.

When classifier $k$ issues a real time detection at date $d_k$, we treat the historical detection errors as an empirical forecast-error sample and approximate their distribution by a normal distribution with mean $\mu_k$ and variance $\sigma_k^2$, exactly as in \citet{michaillat2025}. Letting
\[
\varepsilon_k = d_k - s,
\]
where $s$ is the unknown start date of the current recession, the event $\{s \le t\}$ is equivalent to $\{\varepsilon_k \ge d_k - t\}$. Therefore,
\[
\Pr(s \le t \mid d_k) = \Pr(\varepsilon_k \ge d_k - t).
\]
This probability is computed using only information from the training sample.\\

Whenever an individual recession indicator crosses its threshold, we infer the probability that the recession has already started from the distribution of the detection error. If classifier $k$ is on average exactly on time and the detection error is symmetric, then the recession has started with probability $0.5$ when the classifier is activated. If the classifier tends to detect recessions early on average, this probability is below $0.5$; if the classifier tends to detect recessions late on average, it is above $0.5$. In the months following detection, this probability converges to one according to the cumulative distribution function of the detection error.

More formally, suppose that classifier $k$ detects the $J$ recessions in the training sample. Each detection $j$ yields a detection date $d(k,j)$ and a detection error $\varepsilon(k,j)$. As in \citet{michaillat2025}, we assume that the detection error $\varepsilon(k)$ is normally distributed with mean $\mu(k)$ and standard deviation $\sigma(k)$:
\[
\varepsilon(k) \sim \mathcal{N}\bigl(\mu(k), \sigma^2(k)\bigr).
\]
Then, for any date $t \ge d(k)$, the probability that the start date $s$ of the new recession occurred before $t$, conditional on classifier $k$ having detected a recession at date $d(k)$, is
\[
P(k,t)
= \Pr\bigl(s < t \mid d(k), \mu(k), \sigma(k)\bigr)
= \Pr\bigl(d(k)-s > d(k)-t \mid \mu(k), \sigma(k)\bigr)
= \Pr\bigl(\varepsilon(k) > d(k)-t \mid \mu(k), \sigma(k)\bigr).
\]
Under the normal approximation, this becomes
\[
P(k,t)
= 1 - \Phi\!\left(\frac{d(k)-t-\mu(k)}{\sigma(k)}\right),
\]
where $\Phi(\cdot)$ denotes the cumulative distribution function of the standard normal distribution.

Lastly, we summarize the information from the $K=2$ detection classifiers by constructing an ensemble recession probability. Each classifier produces a probability that a recession has started, and we average these probabilities to obtain a single aggregate measure. Figure ~\ref{fig:recession_probability} shows the resulting ensemble probability for the 1947--2021 training sample. For all 12 recessions, the probability rises in advance of the official start date, reaches 1 at the onset of the recession, and remains there until the recession has ended.

\subsection{Application of Recession Detection to Current Data}

Next, we apply the classifier estimated on the 1947--2021 training sample to the out-of-sample period 2022--2025 to assess recession risk in the United States economy. Figure~\ref{fig:figure_outsample} reports the estimated recession probability over this period. The classifier assigns essentially zero recession probability throughout 2022--2025, indicating no recession signal in the out-of-sample period.


\begin{figure}[t!]
    \centering
    \includegraphics[width=0.8\linewidth]{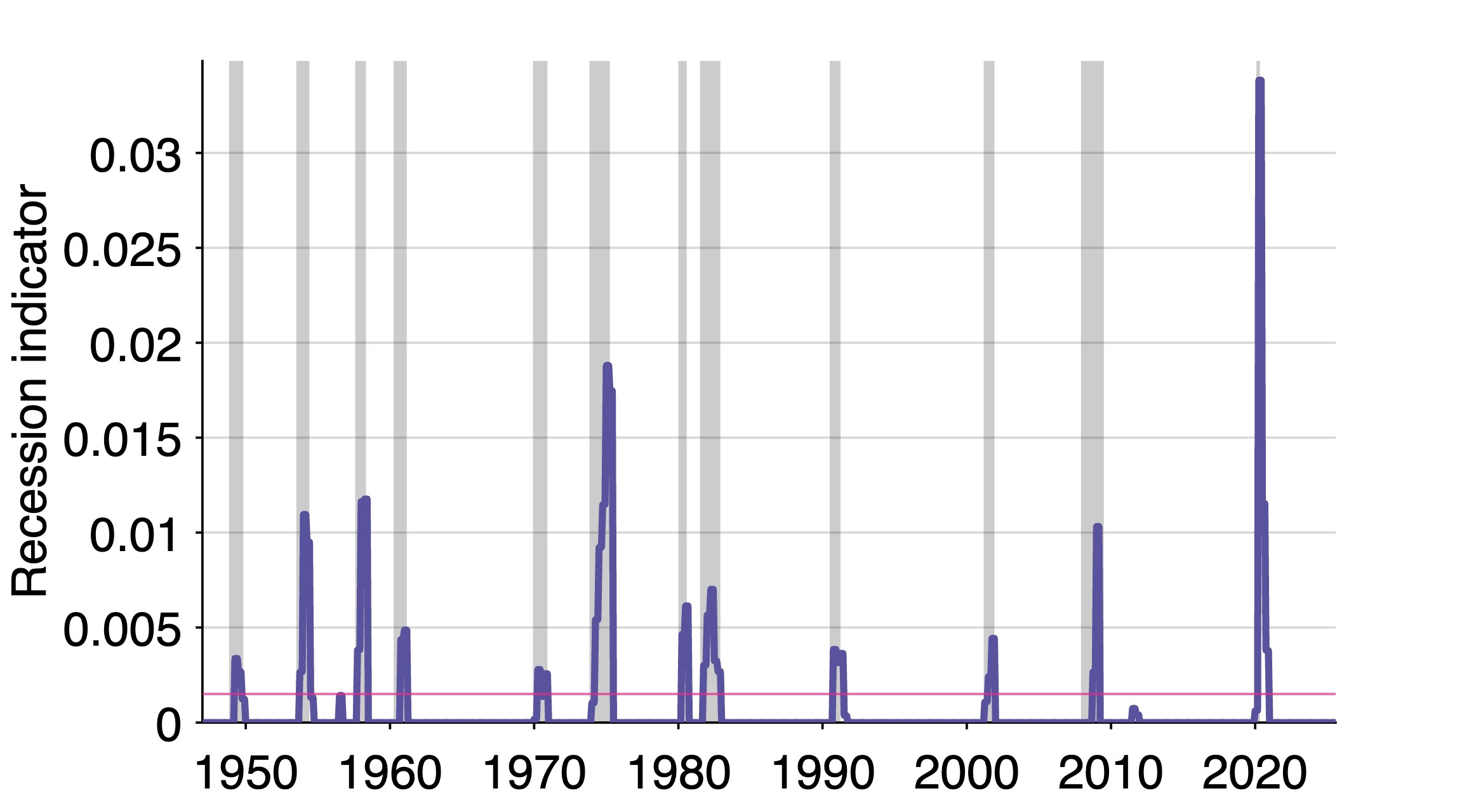}
    \caption{Classifier 1: Threshold 0.0015}
    \label{fig:ensemble1}

    \vspace{0.3em}
    \begin{minipage}{0.95\linewidth}
    \footnotesize
    \textit{Notes:} The classifier specification corresponds to the first classifier reported in Table~\ref{tab:classifier-ensemble}. The purple line represents the recession indicator underlying the classifier, while the pink line represents the corresponding recession threshold. Gray shaded regions indicate recessions dated by the NBER.
    \end{minipage}
\end{figure}

\begin{figure}[H]
    \centering
    \includegraphics[width=0.8\linewidth]{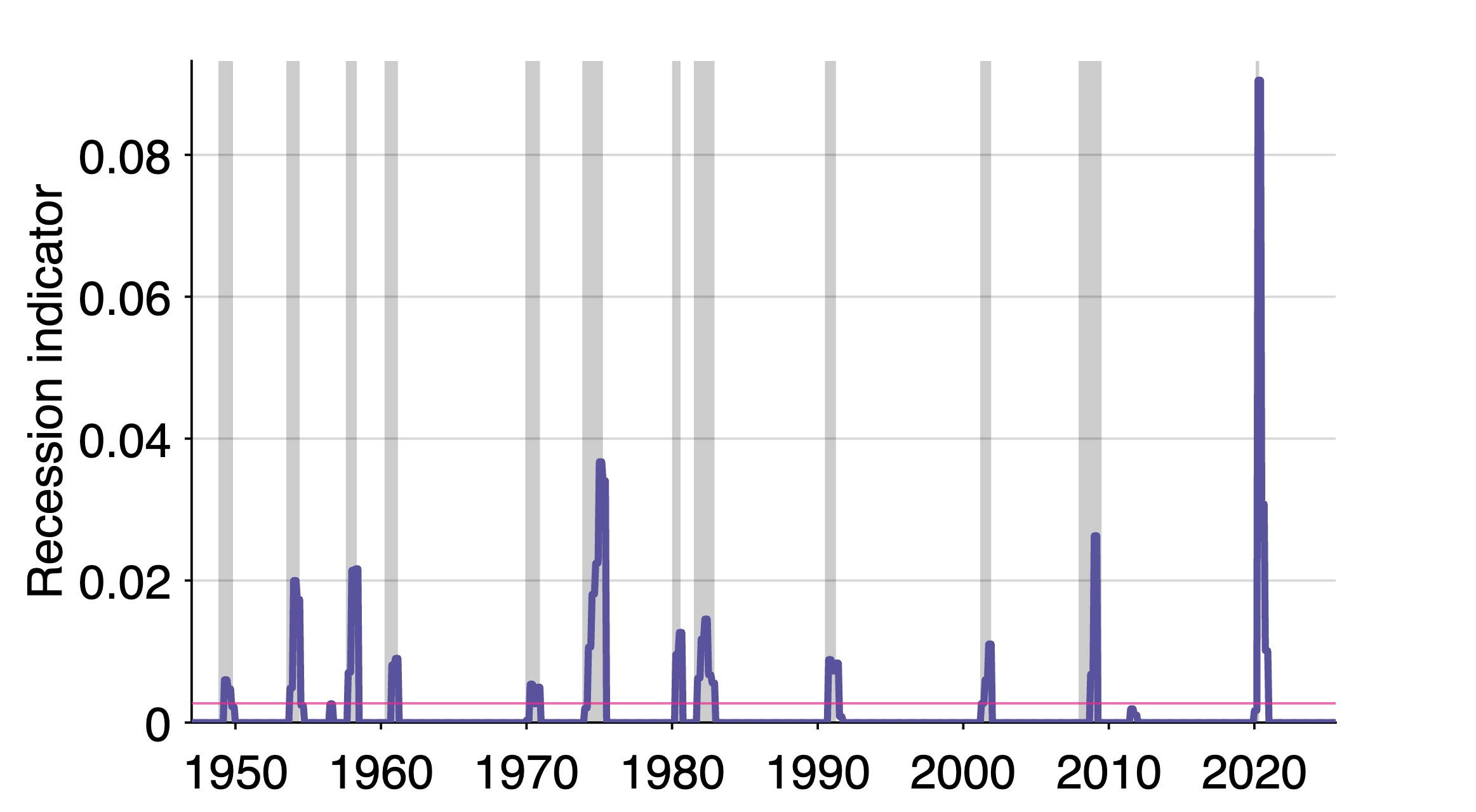}
    \caption{Classifier 2: Threshold 0.0027}
    \label{fig:ensemble2}

    \vspace{0.3em}
    \begin{minipage}{0.95\linewidth}
    \footnotesize
    \textit{Notes:} The classifier specification corresponds to the second classifier reported in Table~\ref{tab:classifier-ensemble}. The purple line represents the recession indicator underlying the classifier, while the pink line represents the corresponding recession threshold. Gray shaded regions indicate recessions dated by the NBER.
    \end{minipage}
\end{figure}

\begin{figure}[H]
\centering
\includegraphics[width=0.8\linewidth]{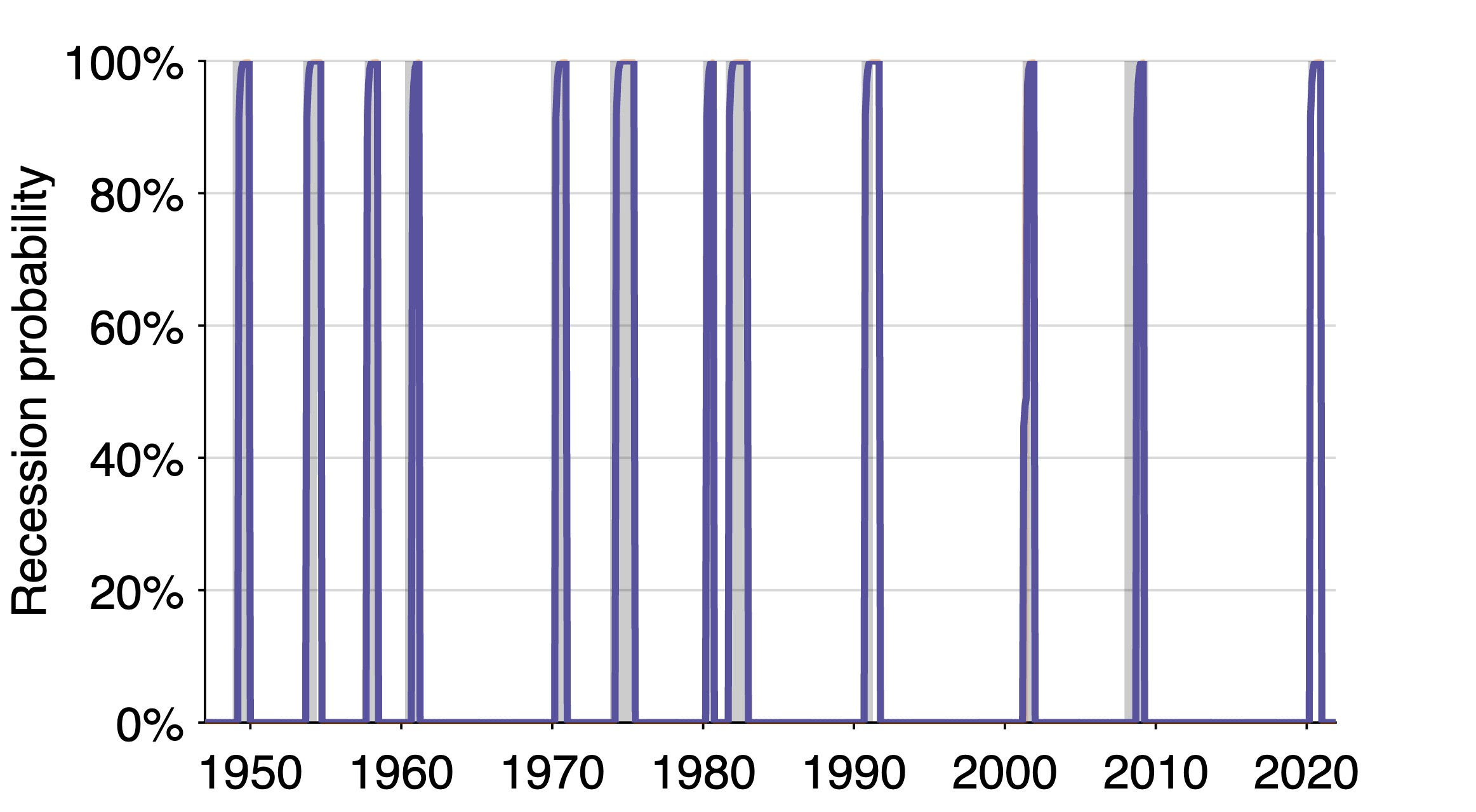}
\caption{A. In-sample Recession Probability, January 1947--December 2021.}
\label{fig:recession_probability}

\vspace{0.3em}
\begin{minipage}{0.8\linewidth}
\footnotesize
\textit{Notes:} The figure plots the estimated probability of a recession over time. The thick purple line shows the average probability across the two classifiers in the selected ensemble. The thin orange lines show the probability from each classifier separately. Shaded areas indicate recession periods.
\end{minipage}
\end{figure}

\begin{figure}[H]
    \centering
    \includegraphics[width=0.8\linewidth]{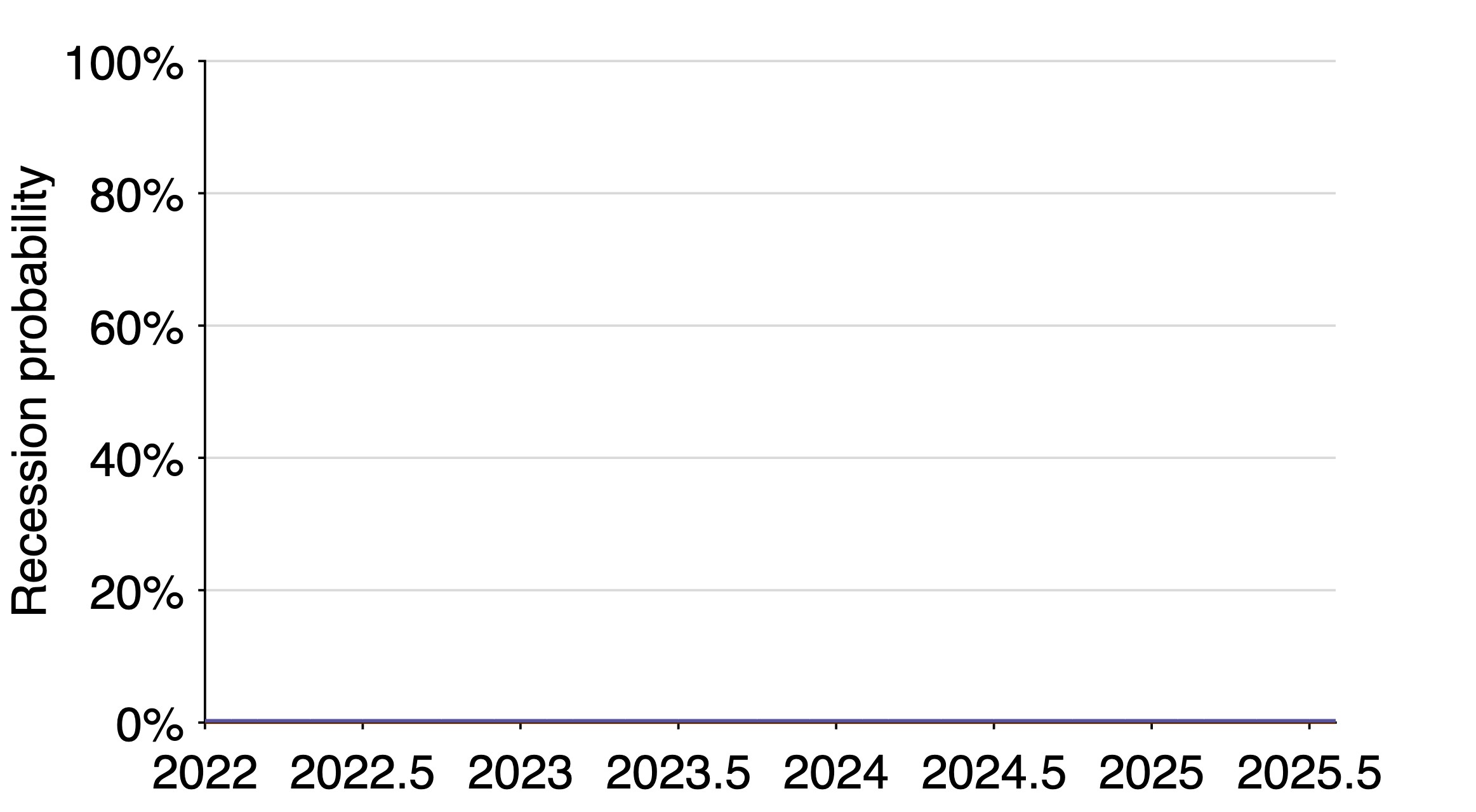}
    \caption{Out-of-sample Recession Probability, January 2022--December 2025}
    \label{fig:figure_outsample}
\end{figure}

\section{Evaluating the recession detection algorithm with backtests}
To examine the reliability of the algorithm, we perform a series of out-of-sample backtests. In each exercise, the training sample is shortened, classifiers are selected using only information available within that training window, and performance is evaluated on recessions occurring after the training period. The results show that the algorithm remains broadly robust across several training windows. However, performance deteriorates when the training sample ends in 2005, suggesting that recession dynamics embedded in historical GDP vintages may not remain fully stable across subsequent business-cycle episodes. We therefore stop the back-testing exercise at the 2005 window.

The full-sample result shows that real time GDP announcements contain enough information to recover historical turning points. The out-of-sample exercise is more demanding as it asks whether the same classifier can be selected before later recession episodes are observed. The weaker performance therefore reflects uncertainty in classifier selection across changing business-cycle environments, rather than a lack of recession information in GDP itself.

\begin{table}[htbp]
\centering
\caption{Out-of-Sample Backtesting Results}
\label{tab:backtest_gdp}
\begin{tabular}{lcc}
\toprule
\textbf{Metric} & \textbf{2015}\\
\midrule
Number of classifiers & 5 \\

\addlinespace
\textit{Training sample detection error (months)} & & \\
Mean & 4.7\\
Standard deviation & 2.0  \\
Minimum & 2.2  \\
Maximum & 9.4 \\

\addlinespace
\textit{Testing sample detection error (months)} & & \\
Mean & 1.0 \\
Standard deviation & 0.0  \\
Minimum & 1.0  \\
Maximum & 1.0 \\

\addlinespace
False positives & 0 \\
False negatives & 0 \\
Training recessions & 11  \\
Testing recessions & 1  \\

\bottomrule
\end{tabular}
\vspace{0.6em}
\begin{minipage}{0.92\linewidth}
\footnotesize
\textit{Note:} The table reports the performance of the selected classifier ensembles in each backtest exercise. 
The training sample is Jan 1947--Dec 2014. The testing period is January 2015--December 2025. Detection errors are measured in months relative to recession onset dates.
\end{minipage}

\end{table}

\subsection{Backtesting from 2015}

We train the algorithm using data through December 2014. The training sample contains 11 recessions, as summarized in Table \ref{tab:backtest_gdp}. This procedure selects 5 classifiers from the high-precision segment of the anticipation–precision frontier, reported in Table \ref{tab:ensemble2015}. We then evaluate these classifiers out of sample using data from January 2015 through December 2025. During this test period, all 5 classifiers detect the 2020 recession with no false positives. The detection is also timely. On average, the recession is signaled only 1 month after its official onset.

\begin{figure}[H]
    \centering
    \includegraphics[width=0.8\linewidth]{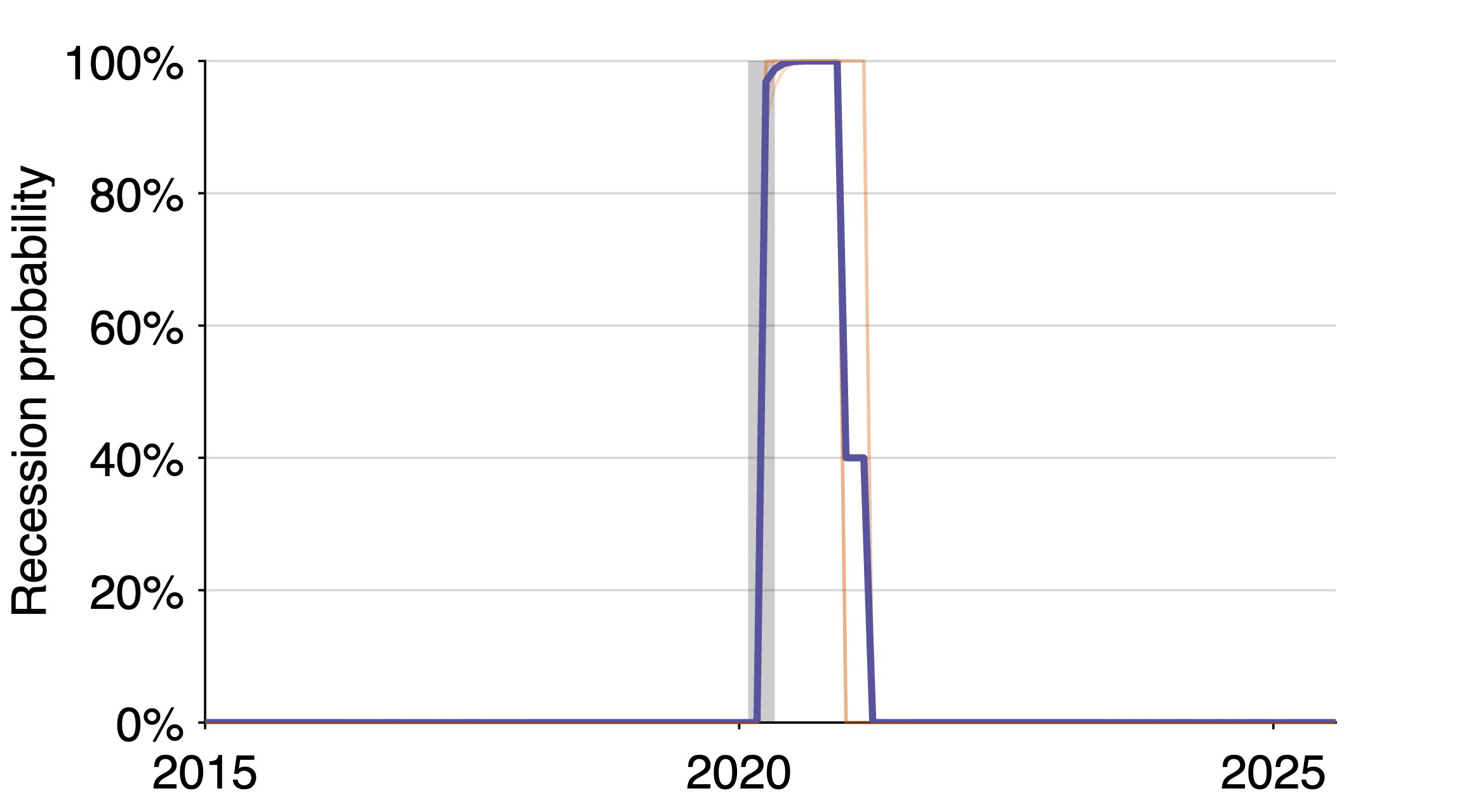}
    \caption{Out-of-sample U.S. Recession Probability, January 2015--December 2025}
    \label{fig:placeholder}
\end{figure}

\subsection{Backtesting from 2005}

We train the algorithm using data through December 2004. The training sample contains ten recessions. This procedure selects one classifier from the high-precision segment of the anticipation–precision frontier, reported in Table \ref{tab:ensemble2005}. We then evaluate the selected classifier out of sample using data from January 2005 through December 2025.

The out-of-sample performance deteriorates relative to earlier backtesting exercises. Although the classifier exhibits anticipatory detection, with a mean detection error of approximately $-5.5$ months, it also generates four false positives during the testing period. This implies that the classifier signals recessions nearly five months before the official recession onset dates on average, but at the cost of reduced precision. The standard deviation of the detection error is approximately 3.5 months, which remains relatively moderate given the difficulty of identifying business-cycle turning points in real time. Nevertheless, the increase in false positives suggests that classifiers calibrated on earlier postwar recession dynamics may not generalize fully to later business-cycle episodes. We therefore stop the backtesting exercise at the 2005 window.

\begin{figure}[H]
    \centering
    \includegraphics[width=0.8\linewidth]{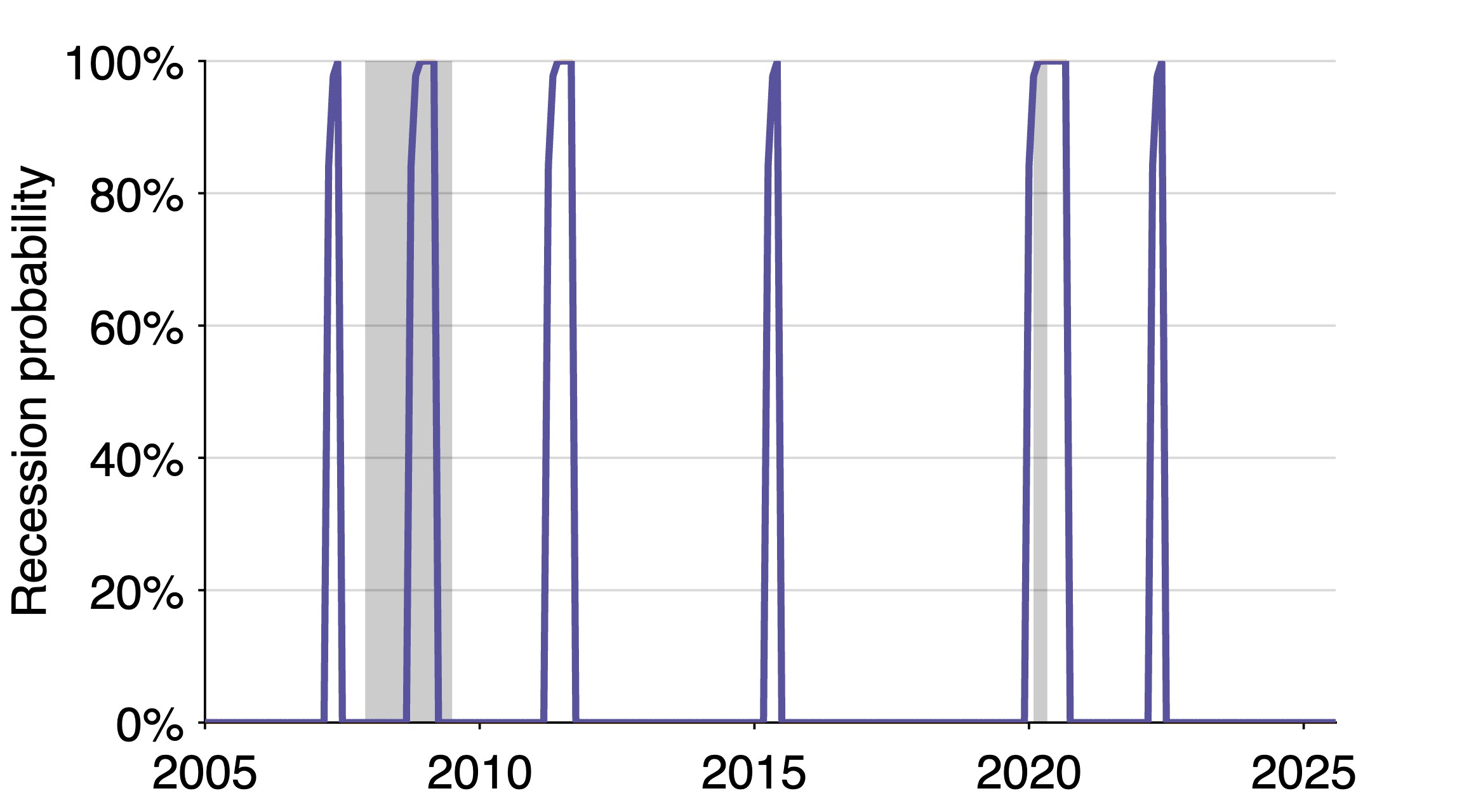}
    \caption{Out-of-sample U.S. Recession Probability, January 2005--December 2025}
    \label{fig:placeholder}
\end{figure}


\section{Conclusion}
A natural question is why one should rely on real time GDP rather than labor-market indicators to proxy business-cycle turning points. Labor-market indicators are often timely and informative, particularly when employment conditions deteriorate before output declines are fully recorded. However, GDP remains the broadest summary measure of aggregate economic activity and is closely tied to the concept of business-cycle turning points used in official recession dating. For this reason, evaluating whether real time GDP announcements alone can approximate NBER-style recession dates provides a useful benchmark. The purpose is not to replace labor-market indicators, but rather to examine whether real time GDP itself contains sufficient information to systematically identify turning points in economic activity.

The results in this paper suggest that real time GDP announcements contain substantial information about business-cycle turning points and can closely approximate official recession dates using only information available at each point in time. While the selected classifiers reproduce historical recession episodes with a high degree of accuracy in sample, the weaker out-of-sample performance indicates that the relationship between GDP dynamics and recession timing may vary across business-cycle episodes. Nevertheless, the findings demonstrate that real time GDP can still provide a transparent and operational framework for real time recession dating, particularly in economies where official business-cycle chronologies are unavailable or announced with substantial delays.

\bibliography{\bib}

\appendix

\section{Additional results}
\begin{figure}[H]
    \centering
    \includegraphics[width=1\linewidth]{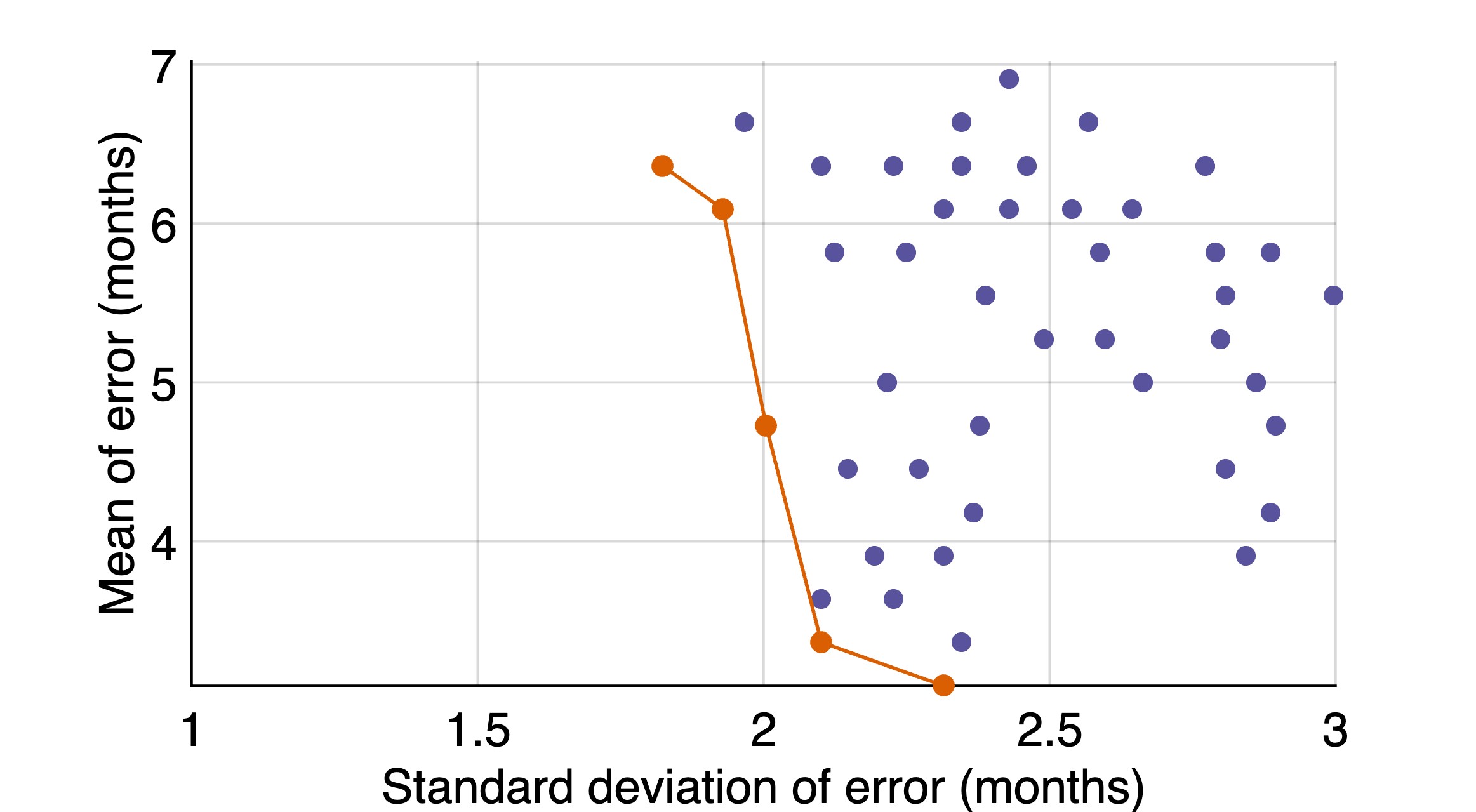}
    \caption{Anticipation Precision Frontier 2015}
    \label{fig:placeholder}
\end{figure}

\begin{figure}[H]
    \centering
    \includegraphics[width=1\linewidth]{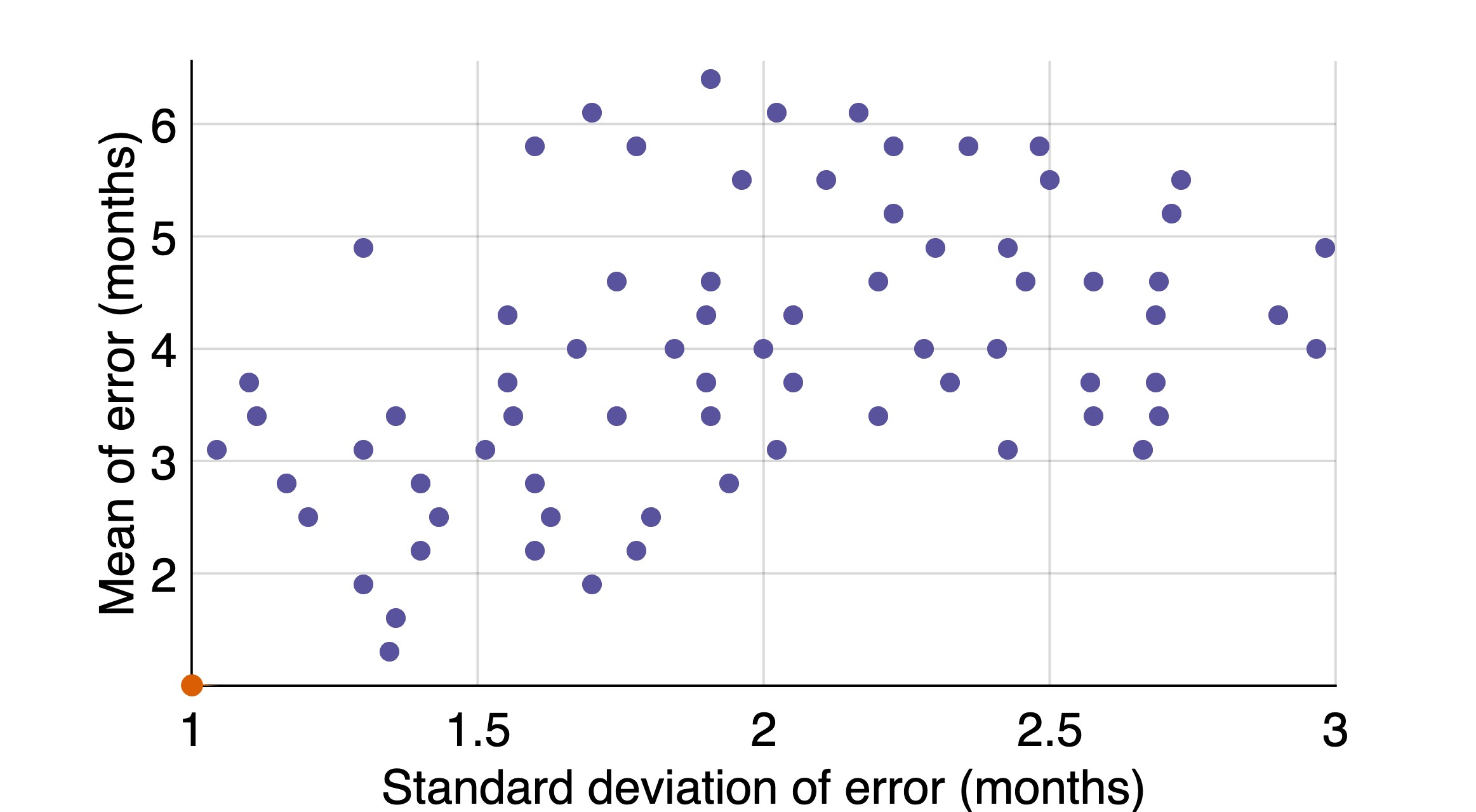}
    \caption{Anticipation Precision Frontier 2005}
    \label{fig:placeholder}
\end{figure}

\begin{table}[h]
\centering
\caption{Classifier Ensemble Selected in the 2015 Backtest}
\label{tab:ensemble2015}
\begin{tabular}{ccccccccc}
\toprule
 & \multicolumn{2}{c}{Smooth} & Curving & Turning & Threshold & Standard & Mean \\
\cmidrule(lr){2-3}
Classifier & Method & Param & Method & Param & Value & Error & Error \\
\midrule
(1) & SMA & 3 & 0.1 & 6 & 0.0039 & 1.823 & 6.364 \\
(2) & SMA & 3 & 0.0 & 6 & 0.0014 & 1.928 & 6.091 \\
(3) & SMA & 2 & 0.2 & 6 & 0.0131 & 2.004 & 4.727 \\
(4) & SMA & 2 & 0.0 & 3 & 0.0015 & 2.101 & 3.364 \\
(5) & SMA & 2 & 0.1 & 3 & 0.0027 & 2.314 & 3.091 \\
\bottomrule
\end{tabular}
\end{table}

\begin{table}[h]
\centering
\caption{Classifier Ensemble Selected in the 2005 Backtest}
\label{tab:ensemble2005}
\begin{tabular}{ccccccccc}
\toprule
 & \multicolumn{2}{c}{Smooth} & Curving & Turning & Threshold & Standard & Mean \\
\cmidrule(lr){2-3}
Classifier & Method & Param & Method & Param & Value & Error & Error \\
\midrule
(1) & SMA & 1 & 0.0 & 6 & 0.0009 & 1.000 & 1.000 \\
\bottomrule
\end{tabular}
\end{table}

\begin{figure}[t!]
    \centering
    \includegraphics[width=1\linewidth]{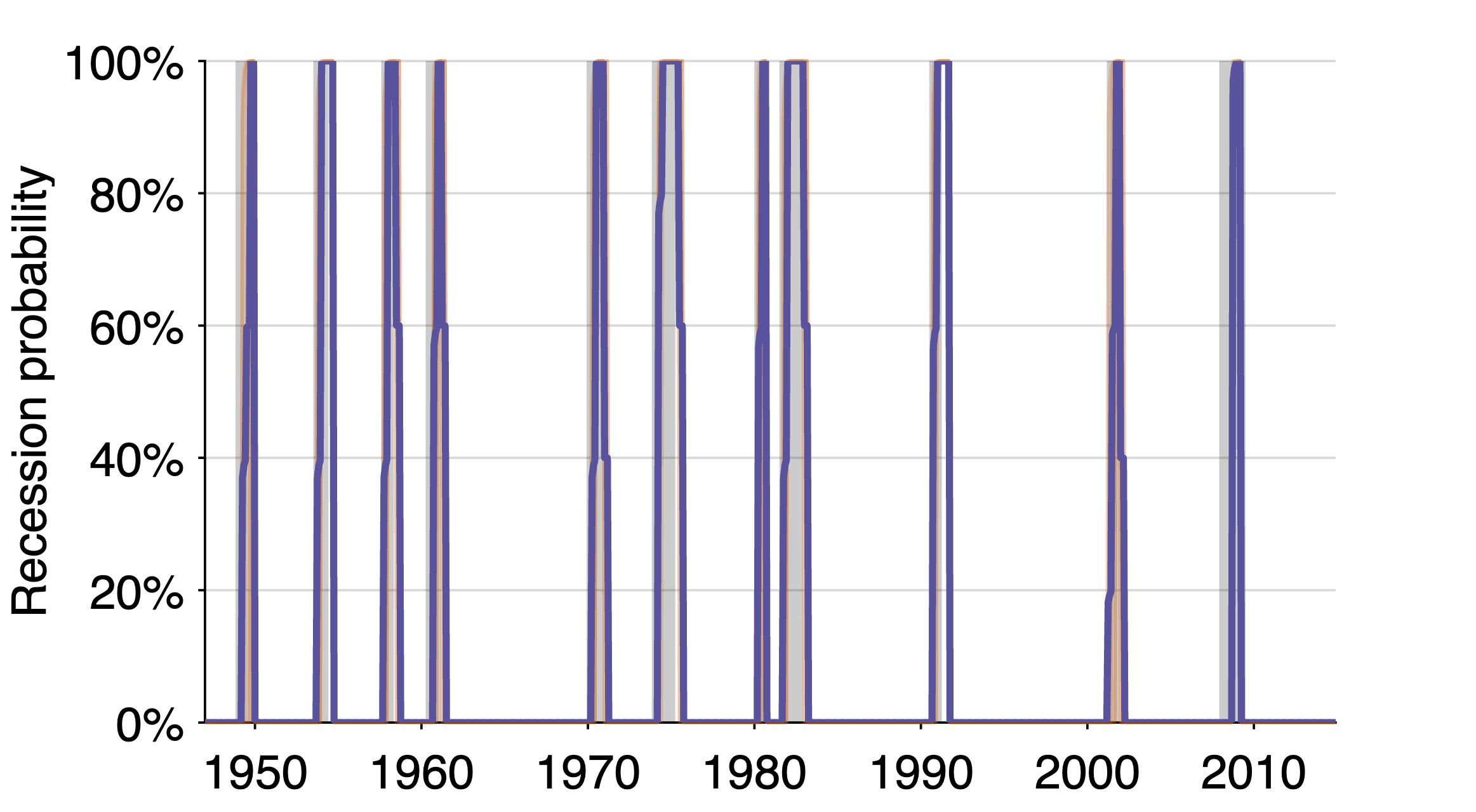}
    \caption{In Sample Recession Probability, January 1947 - December 2014}
    \label{fig:placeholder}
\end{figure}

\begin{figure}[t!]
    \centering
    \includegraphics[width=1\linewidth]{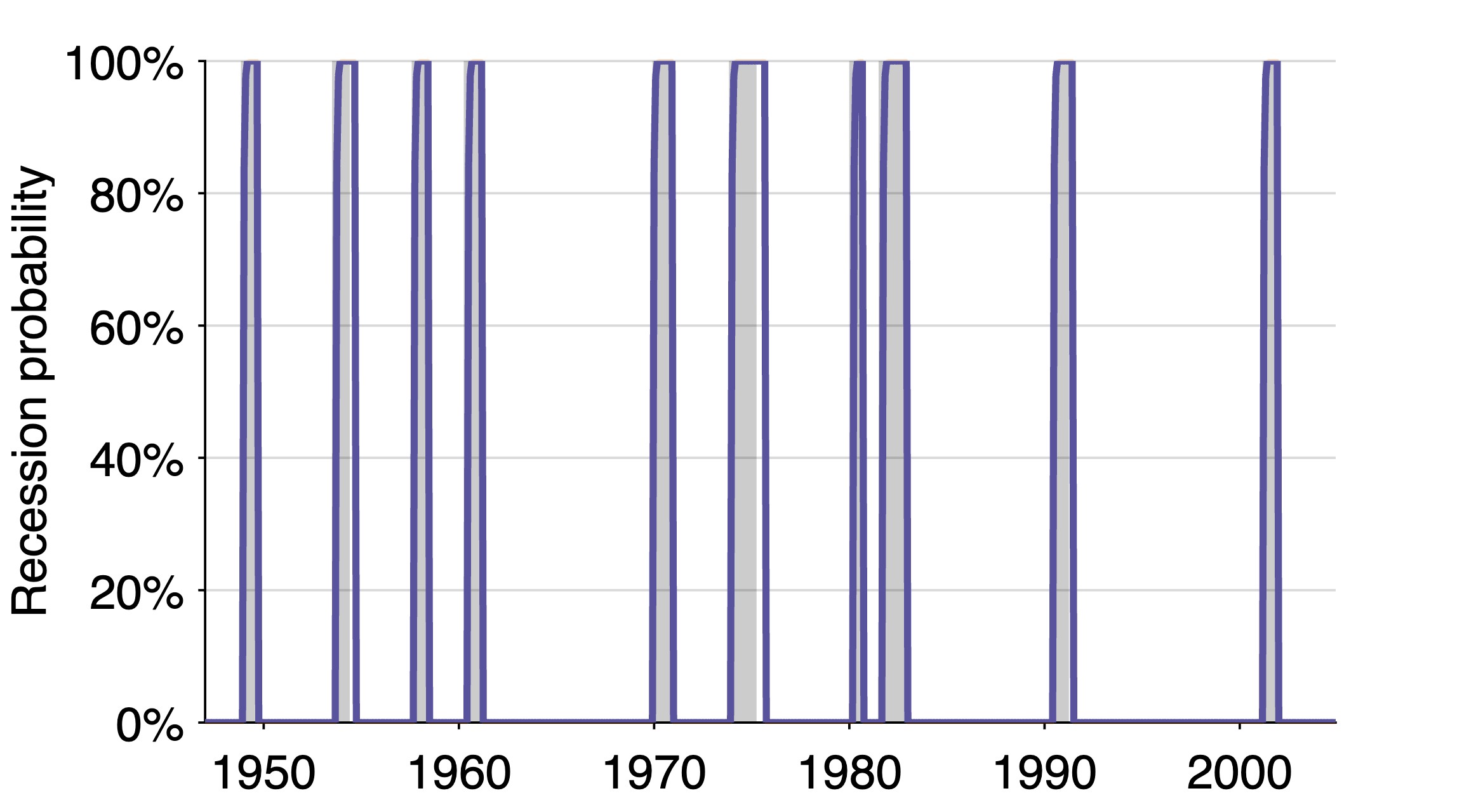}
    \caption{In Sample Recession Probability, January 1947 - December 2004}
    \label{fig:placeholder}
\end{figure}

\end{document}